\documentclass{ws-rv9x6}
\usepackage[square]{ws-rv-van}

\def\bE{\mathbf{ E}}
\def\bv{\mathbf{ v}}
\def\bq{\mathbf{ q}}

\def\bJ{\mathbf{ J}}

\def\bn{\mathbf{ n}}

\begin{document}
\chapter[Thermostated Boltzmann Equation in a Field]{Nonequilibrium Stationary Solutions of Thermostated Boltzmann Equation in a Field} 

\author[F. Bonetto, J.L. Lebowitz]{F. Bonetto\footnote{School of Mathematics, GaTech, Atlanta GA 30332}, J.L. Lebowitz\footnote{Departments of Mathematics and Physics, Rutgers University, Piscataway NJ 08854}}

\begin{dedication}
 Dedicated to Leopoldo Garc\'\i a Col\'\i n on the occasion of his eightieth birthday.
\end{dedication}

\begin{abstract}
We consider a system of particles subjected to a uniform external force $\bE$ and undergoing random collisions with ``virtual'' fixed obstacles, as in the Drude model of conductivity\cite{AM}. The system is maintained in a nonequilibrium stationary state by a Gaussian thermostat. In a suitable limit the system is described by a self consistent Boltzmann equation for the one particle distribution function $f$. We find that after a long time $f(\bv,t)$ approaches a stationary velocity distribution $f(\bv)$ which vanishes for large speeds, {\it i.e.} $f(\bv)=0$ for $|\bv|>v_{max}(\bE)$, with $v_{max}(\bE)\simeq |\bE|^{-1}$ as $|\bE|\to0$. In that limit $f(\bv)\simeq \exp(-c|\bv|^3)$ for fixed $\bv$, where $c$ depends on mean free path of the particle. $f(\bv)$ is computed explicitly in one dimension.
\end{abstract}

\body

\section{Introduction}

Nonequilibrium stationary states (NESS) in real systems must be maintained by interaction with (effectively infinite) external reservoirs. Since a complete microscopic description of such reservoirs is generally not feasible it is necessary to represent them by some type of modeling\cite{Ru}.

However, unlike systems in equilibrium, which maintain themselves without external inputs and for which one can prove (when not inside a coexistence region of the phase diagram) that bulk behavior is independent of the nature of the boundary interactions, we do not know how different microscopic modeling of external inputs, affects the resulting NESS.

One particular type of modeling dynamics leading to NESS is via Gaussian thermostats, see \cite{BL1}. Analytical results as well as computer simulations have shown that the stationary states produced by these dynamics behave in many cases in accord with those obtained from more realistic models\cite{Ev}.
This has led us to continue our study of the NESS in current carrying thermostated systems. In our previous work, see \cite{BL1,BL2} we carried out extensive numerical and analytical investigations of the dependence of the current on the electric field for a model system consisting of $N$ particles with unit mass, moving among a fixed periodic array of discs in a two dimensional square $\Lambda$ with periodic boundary conditions, see Fig. 1. They are acted on by an external (electric) field $\bE$ parallel to the $x$-axis and by a ``Gaussian thermostat''.  (The discs are located so that there is a finite horizon, {\it i.e.} there is a maximum distance a particle can move before hitting a disc or obstacle).

\begin{figure}
\centerline{\psfig{file=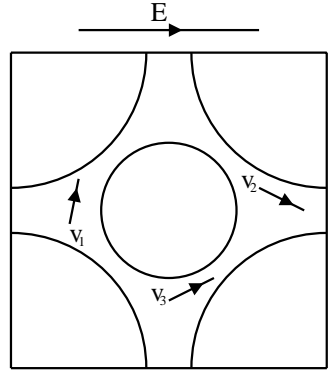,width=6cm}}
\caption{General billiard structure with three particles shown.}
\label{fig1.1}
\end{figure}

The equations of motion describing the time evolution of the positions $\bq_i$ and velocities $\mathbf{ v}_i$, $i=1,...,N$, are:

\begin{eqnarray}\label{dyn1}
\dot\bq_i&=&\bv_i\qquad
\bq_i=(q_{i,x},q_{i,y})\in\Lambda'\\
\dot\bv_i&=&\bE-\alpha(\bJ,U)\bv_i+F_{obs}(\bq_i)
\end{eqnarray}
where

\begin{equation}\label{dyn3}
\alpha(\bJ,U)=\frac{\bJ\cdot\bE}{U},
\qquad\bJ=\frac{1}{N}\sum_{i=1}^N\bv_i, \qquad U=\frac{1}{N}\sum_{i=1}^N\bv_i^2
\end{equation}
Here $\Lambda'=\Lambda\backslash{\cal D}$, with ${\cal D}$ the region occupied by the discs (obstacles) and $F_{obs}$ represents the elastic scattering which takes place at the surface of the obstacles. The purpose of the Gaussian thermostat, represented by the term $\alpha(\bJ,U)\mathbf{ v}$ in \eref{dyn1}, is to maintain the total kinetic energy $1/2\sum_{i=1}^N\bv_i^2$ constant, {\it i.e.} $U=v_0^2$. It also has the effect of making the flow $\Phi_t$ generated by \eref{dyn1} on the $(4N-1)$ dimensional energy surface non Hamiltonian when $\mathbf{ E}\not=0$. Another effect of the thermostat
is to effectively couple all the particles in a mean field way, $\alpha(\mathbf{ J},U)$, depending only on the total momentum of the particles.  Note that this is the only coupling between the particles
in this system.

Our main interest is in the NESS of this model system. To get some analytical handle on the form of the NESS we also investigated  numerically a model system in which the deterministic collisions with the obstacles are replaced by a stochastic process in which particle velocities get their orientations changed at random times, independently for each particle \cite{BL2}.  This yields, in the limit $N\to\infty$, a self consistent Boltzmann equation. In the spatially uniform case, the only one we shall consider here, the equation takes the form

\begin{equation}\label{be}
\frac{\partial}{\partial t} f(\bv,t)+\frac{\partial}{\partial \bv}\left[(\bE-\mu \bv)f(\bv,t)\right]=-\lambda \int_{\bv\cdot \bn<0}\frac{\bv'\cdot \bn}{2}\left(f(\bv,t)-f(\bv',t)\right)d\bn
\end{equation}
where $\bv'=\bv-2\bn(\bn\cdot \bv)$, $\bn$ is a unit vector and $\mu$ is determined self consistently
by requiring that

\begin{equation}
 \frac{d}{dt}\int \frac{\bv^2}{2} f(\bv,t)d\bv=-\bE\cdot \langle \bv\rangle+\mu \langle \bv^2\rangle=0
\end{equation}
or
\begin{equation}\label{self}
 \mu=\frac{\bE\cdot\langle \bv\rangle}{\langle \bv^2\rangle}
\end{equation}
This ensures that $\langle \bv^2\rangle$ is independent of time, {\it i.e.} $\int\bv^2 f(\bv,t)d\bv=\int\bv^2 f(\bv,0)d\bv$. When $\bE=0$ also $\mu=0$ and $f(\bv,t)$ will approach, as $t\to\infty$, the function $\bar f(|\bv|,0)=\frac{1}{S(1)}\int f(\bv,0)d\bn$, where $S(1)$ is the area of the unit sphere in $n$ dimensions. There will thus not be a unique stationary solution of \eref{be}. The situation is different when $\bE\not=0$. In that case $\lim_{t\to\infty} f(\bv,t)=f(\bv)$ independent of $f(\bv,0)$. It is this NESS of \eref{be} which we shall study here.

The outline of the rest of the paper is as follows. In section 2 we show that the stationary solution vanishes for $|\bv|>|\bE|/\mu$. In section 3 we obtain an exact expression for the steady state solution of this equation in one dimension. In section 4 we investigate the small field limit of this stationary distribution. Finally in section 5 we find the small field limit of the distribution in two dimensions.

\section{Properties of \eref{be} in arbitrary dimension}

Here we investigate some general properties of \eref{be} in the steady state. Observe that the right hand side of the equation preserves $\bv^2$. On the other hand we have

\begin{equation}
 (\bE-\mu \bv)\cdot \bv\leq 0\qquad \hbox{if }\quad |\bv|>\frac{|\bE|}{\mu}
\end{equation}
Thus if we define 

\begin{equation}
F(V,t)=\int_{|\bv|\leq V}f(\bv,t)d\bv
\end{equation}
we get that

\begin{eqnarray}
 \frac{d}{dt}F(V,t)&=&-\int_{|\bv|=V}([(\bE\cdot \bv)-\mu V^2]f(\bv,t)d\bv\leq\\
&\leq& -{\cal S}(V)(|\bE|-\mu|V|)\sup_{|\bv|=V}f(\bv,t)
\end{eqnarray}
where ${\cal S}(V)$ is the surface of the sphere of radius $V$. Since in the steady state we must have $\dot F(V,t)=0$ it follows that $f(\bv)=0$ if $\bv>|\bE|/\mu$, where $\mu$ is now the stationary self consistent value. When $|\bE|\to 0$, $\mu$ will decrease as $|\bE|^2$, as long as $\lambda>0$, and so the maximum speed will grow as $|\bE|^{-1}$.

Observe that, as noted before, the limit for $\bE\to 0$ is very different from that obtained when $\bE=0$ and $\mu=0$. This behavior is due to the fact that the collision term preserves $|\bv|^2$. In particular it would fail if one also adds a non linear Boltzmann particle-particle collisions term to the right hand side of \eref{be}. In that case the stationary distribution would be positive for all $\bv$. We expect however that \eref{be} would describe the behavior of the system for a long time when the effective rate of inter-particle collisions is small compared to the rate of collisions with the obstacles. 

\section{The Boltzmann Equation in one dimension}

We consider the one dimensional version of \eref{be}. This takes the form:

\begin{equation}\label{be1}
 \frac{\partial}{\partial t} f(v,t)+\frac{\partial}{\partial v}\left[(E-\mu v)f(v,t)\right]=-\lambda |v|\left(f(v,t)-f(-v,t)\right)
\end{equation}
with $\mu$ determined self consistently by the requirement that the average energy remain fixed, see \eref{self}.  We look for a stationary solution of \eref{be} in the form

\begin{equation}
 f(v)=\phi(v)+E\psi(v)
\end{equation}
with

\begin{eqnarray}\label{sym}
 \phi(v)&=&\phi(-v)\crcr
 \psi(-v)&=&-\psi(-v)
\end{eqnarray}
and $\phi$ and $\psi$ defined for $v>0$. The time independent \eref{be1} takes the form

\begin{eqnarray}\label{due}
 \frac{\partial}{\partial v}\left[\mu v\phi(v)-E^2\psi(v)\right]&=&0\crcr
 \frac{\partial}{\partial v}\left[\phi(v)-\mu v\psi(v)\right]&=&-2\lambda v \psi(v)
\end{eqnarray}
where $\mu$ now has its steady states value given by \eref{self}. Using the fact that $\psi$ is odd the first equation implies that

\begin{equation}
 \psi(v)=\frac{\mu v}{E^2}\phi(v).
\end{equation}
Setting $\mu=\nu E^2$ we have $\psi(v)=\nu v\phi(v)$ and the second equation in \eref{due} becomes

\begin{equation}\label{odd}
 \frac{\partial}{\partial v}\left[(1-\nu^2 E^2 v^2)\phi(v)\right] = -2\lambda \nu v^2 \phi(v)
\end{equation}
Observe that since $f(v)\geq 0$ we have from the definition, \eref{sym}, that $\phi(v)\geq 0$. It follows from \eref{odd} that, if $\phi(v)>0$ for $v>\frac{1}{\nu E}$, then we will have $\lim_{v\to\infty} \phi=\infty$, so we must have 

\begin{equation}
\phi(v)=0 \qquad\hbox{for }\qquad v>\frac{1}{\nu E} 
\end{equation}
This is consistent with the general result in the previous section. Setting

\begin{equation}
 \Phi(v)=(1-\nu^2 E^2 v^2)\phi(v)
\end{equation}
\eref{odd} becomes

\begin{equation}
 \frac {d}{dv}\log \Phi(v)=\frac{\Phi'(v)}{\Phi(v)}=-\frac{2\lambda \nu v^2}{1-\nu^2 E^2 v^2}=\frac{2\lambda}{\nu E^2}-\frac{2\lambda}{\nu E^2}\frac{1}{1-\nu^2 E^2 v^2}
\end{equation}
This can be solved to give

\begin{equation}
 \Phi(v)=C\exp\left(\frac{2\lambda}{\nu E^2}v + \frac{\lambda}{\nu^2 E^3}\log\left(\frac{1-\nu E v}{1+\nu E v}\right)\right)
\end{equation}
or
\begin{equation}\label{Npdis}
 \phi(v)=\frac{C}{1-\nu^2 E^2 v^2}\left(\frac{1-\nu E v}{1+\nu E v}\right)^{\frac{\lambda}{\nu^2 E^3}}e^{\frac{2\lambda}{\nu E^2}v}
\end{equation}
for $v<\frac{1}{\nu E}$ and 0 otherwise. We finally get, using \eref{sym}, that for $v\in\mathbb{R}$,

\begin{equation}\label{fin}
f(v)= \frac{C}{1-\nu E v}\left(\frac{1-\nu E |v|}{1+\nu E |v|}\right)^{\frac{\lambda}{\nu^2 E^3}}e^{\frac{2\lambda}{\nu E^2}|v|}
\end{equation}
for $|v|<\frac{1}{\nu E}$ and 0 otherwise. $C$ is a normalization constant. 

Observe that at $v=(\nu E)^{-1}$ the solution is singular if $\lambda <\nu^2E^3$. This means that the field is strong enough to push the velocity of the particle close to its limiting value before a collision happen\footnote[1]{Note that if we let $\lambda\to 0$ than the singularity becomes non integrable and:
\begin{equation}
 \lim_{\lambda\to 0} \phi(v)=\frac{1}{2}\delta\left(|v|-\frac{1}{\nu E}\right)
\end{equation}
so that
\begin{equation}
 f(v)=\delta\left(v-\frac{1}{\nu E}\right)
\end{equation}
with $\langle v^2\rangle=(1/\nu E)^2=v_0^2$ specified a priori, so that $\nu=1/E v_0$ or $\mu=E/ v_0$. We also have $\langle v \rangle=v_0$. This is exactly what happens in the $N$-particles $d$-dimensional thermostated case without collisions.All particles end up moving parallel to the field with the same speed independent of the strength of $|\bE|>0$.}. Furthermore, when the field goes to 0 we do not obtain a Maxwellian distribution. To be sure, there is no reason why this should happen since we do not have any mechanism (such as inter-particle collisions) which would tend to bring the system to equilibrium, as one would expect to be the case in a realistic physical model.

It follows from \eref{odd} that
\begin{equation}
 \int_0^{\frac{1}{\nu E}}v^2\phi(v)dv=\frac{C}{2\lambda\nu}
\end{equation}
so that the average kinetic energy in the steady state is $\langle v^2\rangle=C/\lambda \nu=v_0^2$ and the average current is $\langle v\rangle=CE/\lambda \nu$.

\section{Small field limit}

\Eref{fin} can be rewritten in the form:

\begin{equation}
 f(v)=\frac{C}{1-\nu E v}\exp\left[{\frac{\lambda}{\nu^2 E^3}}\left(\log\left(1-\nu E |v|\right)-\log\left(1+\nu E |v|\right)\right)\right]e^{\frac{2\lambda}{\nu E^2}|v|}
\end{equation}
This expression is clearly analytic in $E$ for $E<1/\nu |v|$. In particular we can compute the zero order term by expanding $\log\left(1-\nu E |v|\right)-\log\left(1+\nu E |v|\right)$ in term of $\nu E |v|$. Clearly only terms odd in $E$ will appear. The term linear in $E$ cancel out the exponent $\frac{2\lambda}{\nu E^2}|v|$. The term proportional to $E^3$ gives, once multiplied by $\lambda/\nu^2 E^3$, a term $2/3\lambda \nu |v|^3$. This gives us:

\begin{equation}\label{ex}
 f(v)=C(1+E\nu v)\exp\left(-\frac{2}{3}\lambda\nu v^3\right)+o(E)
\end{equation}
where the term linear in $E$ comes from the prefactor $1/(1-\nu E v)$. Here $\nu$ is the limiting value of $\nu$ which, given the (initial) kinetic energy $v_0^2/2$, is given by

\begin{equation}\label{K}
 \nu=\frac{K}{v_0^3\lambda}+O(E^2)
\end{equation}
with 
\begin{equation}
 K=\frac{3}{8}\frac{2^\frac{1}{2}3^\frac{3}{4}\Gamma\left(\frac{2}{3}\right)^\frac{3}{2}}{\pi^\frac{3}{2}}
\end{equation}

We can also obtain the behavior of $f(v)$ when $E$ tends to 0 by setting $E=0$ in \eref{odd}. This gives:

\begin{equation}
 \frac{\partial}{\partial v}\phi(v)=-2\lambda \nu v^2 \phi(v)
\end{equation}
whose solution is found to be
\begin{equation}
\phi(v)=Ce^{-\frac{2}{3}\lambda \nu v^3}
\end{equation}
This agrees with the result in \eref{ex} when $E\to 0$.

\section{The general case for small $\bE$}

In the two dimensional case the stationary version of \eref{be} takes the form

\begin{equation}\label{be2}
\frac{\partial}{\partial \bv}\left[(\bE-\nu|\bE|^2 \bv)f(\bv)\right]=-\lambda \int_{\bv\cdot \bn<0}\frac{\bv'\cdot \bn}{2}\left(f(\bv)-f(\bv')\right)d\bn
\end{equation}
where $\bv'=\bv-2\bn(\bn\cdot \bv)$, $\bn$ is a unit vector and we have set $\mu=\nu|\bE|^2$.

We do not have an exact solution of this equation for arbitrary field strength. We can however still find the small $|\bE|$ behavior of $f(\bv)$ as we did before for the one dimensional case. To do this we first expand $f(\bv)$ in harmonics of the angular part of $\bv$. This means that that we write:

\begin{equation}
 f(\bv)=\sum_{n=0}^\infty f_n(v)\cos(n\theta)
\end{equation}
where $\bv=(v\cos(\theta),v\sin(\theta))$. We have taken the field to be in the $x$-direction, $\bE=(E,0)$, so that no term in $\sin(\theta)$ appears due to the symmetry of the system with respect to the $x$ axis. Substituting in \eref{be2} we obtain:
\begin{eqnarray}\label{expa}
&-&E^2 \nu v\partial_vf_n(v)-2E^2 \nu f_n(v)+\\
&+&E\partial_v\frac{f_{n-1}(v)+f_{n+1}(v)}{2}- E\frac{(n-1)f_{n-1}-(n+1)f_{n+1}}{2v}=\nonumber\\ 
&=&\left(\frac{(-1)^{n+1}}{4n^2-1}-1\right)\lambda v f_n(v) \nonumber
\end{eqnarray}
where we have set $f_{-1}\equiv 0$.
Since we expect that the distribution $f(\bv)$ will depend only on $|\bv|$ when $E\to 0$ we can assume that $f_n(v)=E g_n(v)$ for $n\not = 0$.
When $n=0$, the right hand side of \eref{expa} vanishes. The equation thus becomes, after symplifying a common $E^2$ factor,
\begin{equation}
 \nu v \partial_v f_0(v)+ 2\nu f_0(v)- \frac{1}{2}\partial_v g_1(v)-\frac{1}{2v}g_1(v)=0
\end{equation}
while the leading term in $E$ of the equation for $n=1$ is

\begin{equation}
 \partial_v f_0(v)=-\frac{4}{3}\lambda vg_1(v).
\end{equation}
So that, at first order in $E$, we have:
\begin{equation}\label{2ex}
 f(\bv)=C(1+2\nu \bE\cdot \bv)\exp\left(-\frac{8}{9}\lambda\nu |\bv|^3\right)
\end{equation}
This is the same form as \eref{ex} with the limiting value for $\nu$ given by \eref{K} with

\begin{equation}
 K=\frac{1}{9}\frac{3^\frac{3}{4}2^\frac{1}{2}\pi^\frac{3}{2}}{\Gamma\left(\frac{2}{3}\right)^3}
\end{equation}

\section{Concluding remarks}

A similar analysis shows that the small $|\bE|$ behavior of the NESS described by \eref{be} will be of the same form as that given by \eref{2ex} in all dimension $d$, in particular the zero order term will be an exponential in $-|\bv|^3$. We do not however have any intuitive heuristic explanation for this behavior on the more microscopic many particles level.

Let us go back now to our original problem, the NESS for the $N$ particles system in an electric field and Gaussian thermostat, described by Eqs. (\ref{dyn1})-(\ref{dyn3}). As mentioned in the introduction, we expect that in the limit $N\to\infty$ this NESS will agree with our results for the stationary distribution of the Boltzmann equation (\ref{be}) {\it i.e.} there will be a cut-off in the particle speed independent of $N$ when $N$ becomes very large. This indicates that the NESS for the original model will be, for large $N$, concentrated on those parts of the $dN-1$ dimensional sphere in which the particle speeds are closer to each other than they would for a uniform distribution on the sphere. The latter is of course what gives rise to a Maxwellian distribution with the same mean energy. This is consistent with what we found in \cite{BL2}, see in particular section III there.

\section*{Acknowledgment}

The work of FB was supported in part by NSF grant 0604518. The work of JLL was supported in part by NSF grant DMR08021220 and by AFOSR grant AF-FA9550-07. 

%
%
%

\end{document}